\documentclass[reprint, 10pt,a4paper, aps, prl, hidelinks, floatfix]{revtex4-1}
\usepackage[utf8]{inputenc}
\usepackage{amsmath}
\usepackage{amsfonts}
\usepackage{amssymb}
\usepackage{mathrsfs}
\usepackage[hidelinks]{hyperref}
\usepackage{empheq}
\usepackage{graphicx}

\newcommand{\bv}{\mathbf{v}}

\newcommand{\bq}{\mathbf{q}}

\begin{document}

\begin{abstract}
Stratification due to ion-ion friction in a magnetized multiple-ion species plasma is shown to be accompanied by a heat pump effect, identified for the first time here, transferring heat from one ion species to another as well as from one region of space to another.
The heat pump is produced via newly identified heating mechanisms associated with charge incompressibility and the Ettingshausen effect.
Besides their academic interest, these new effects may have useful applications to plasma technologies that involve rotation or compression. 

\end{abstract}

\title{Heat Pump via Charge Incompressibility in a Collisional Magnetized Multi-Ion Plasma}
\date{\today}
\author{M. E. Mlodik, E. J. Kolmes, I. E. Ochs, and N. J. Fisch}
\affiliation{Department of Astrophysical Sciences, Princeton University, Princeton, New Jersey, USA, 08544 and \\
Princeton Plasma Physics Laboratory, Princeton, New Jersey, USA, 08540}

\maketitle 

\textit{Introduction.} 
A magnetized multiple-ion species quasineutral plasma subjected to a changing potential field evolves on multiple transport timescales. 
The fastest timescale is associated with the ion-ion friction, which does not move net charge across field lines, and results in ion stratification according to mass-to-charge ratio $m/Z$ \cite{Kolmes2018, Braginskii1965, Spitzer1952, Taylor1961ii}. 
Since ion-ion friction does not move net charge across the field lines,
the system reaches force equilibrium without rearranging the total ion charge density. Fluid elements of different ion species can only displace each other. 
Therefore, magnetized low-$\beta$ multi-ion species plasma behavior resembles buoyancy in an incompressible fluid. 
Note, however, that, in a buoyant liquid, locally conserved quantities such as density are tied to a fluid element of that liquid, while, in a plasma, charge conservation is tied to an element of the electron fluid. Therefore, plasma buoyancy still allows compression of the ions if charge states $Z$ are different. 

\begin{figure}
    \centering
    \includegraphics[width=\linewidth]{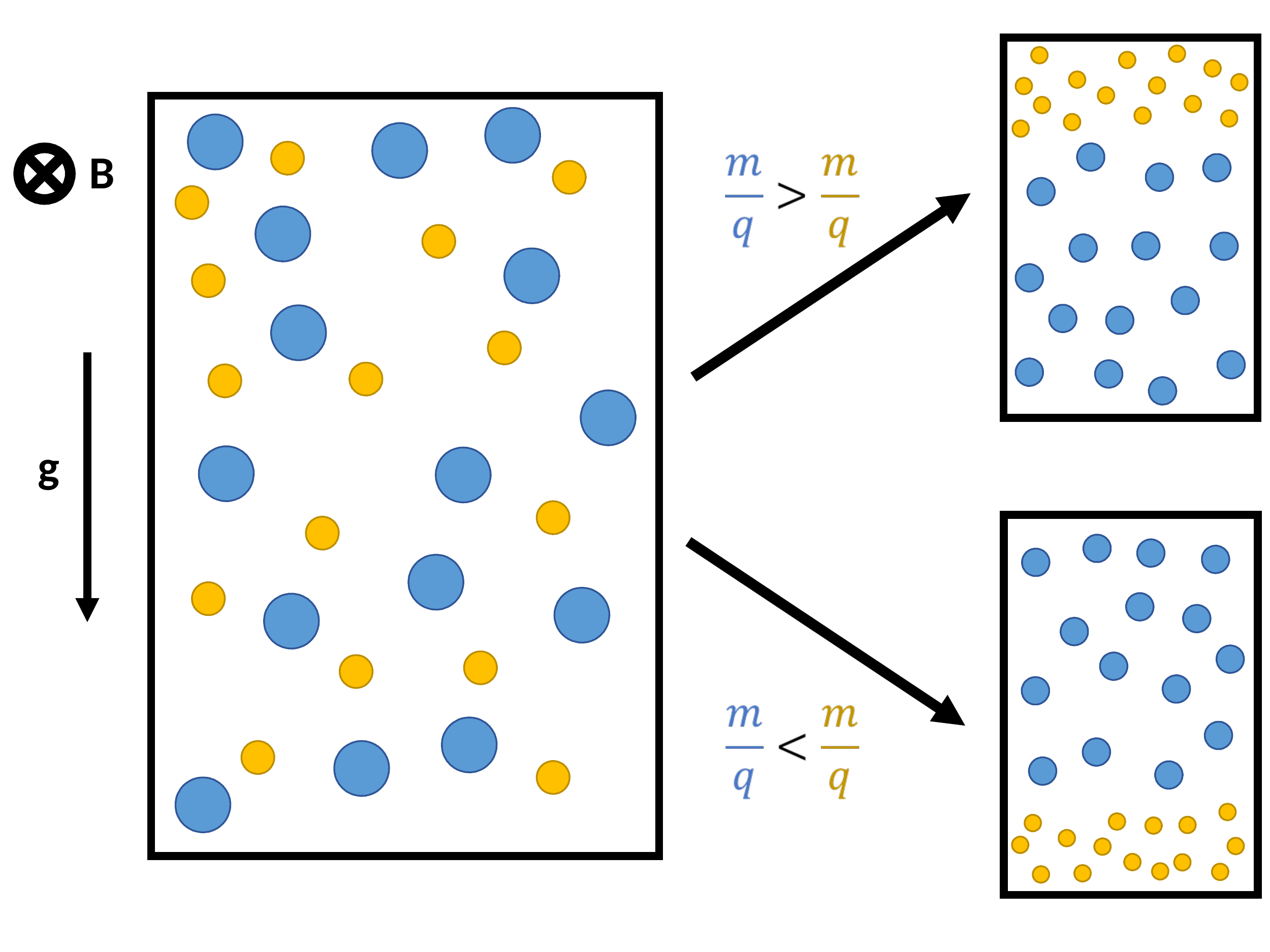}
    \caption{Charge incompressibility in crossed magnetic and gravitational fields. Initially the plasma slab has uniform density and temperature. As external force is applied, the total ion charge density is 
    conserved locally. In contrast, in incompressible fluid, the number density is conserved instead.}
    \label{fig:cartoon}
\end{figure}

This makes all the difference. Magnetized low-$\beta$ multi-ion species plasma is a charge-incompressible substance, possessing very curious thermodynamic properties. The new heat pump effect arises when many low charge state ions try to replace just a few high charge state ions (see Figure~\ref{fig:cartoon}).
To observe charge incompressibility, low-$Z$ ions are compressed and heated, while high-$Z$ ions are decompressed and cooled. 
An additional, but totally separate, heat transport also arises from the Ettingshausen effect \cite{Hinton}, which is a heat flux, Onsager symmetric to the thermal friction. 
In magnetized plasma, cold ions participate in cross-field transport more than hot ions, since they are more collisional. 
Therefore, heat flux appears, and its direction is the opposite to the direction of particle flux. 
Heat flux per particle due to the Ettingshausen effect is larger for light ions. 
Taking charge incompressibility into account, the Ettingshausen effect leads to heat transfer from low-$mZ$ to high-$mZ$ species. 

Remarkably, these heating mechanisms combine to produce a heat pump effect.
If plasma consisted of two ion species, one of them being low-$Z$ and high-$mZ$, and another being high-$Z$ and low-$mZ$, the Ettingshausen effect and heating due to charge incompressibility would act in the same direction. In perhaps more usual cases, they counteract each other. 
Since these mechanisms turn out to have the same scaling and differ only in magnitude, heat is transferred from the region where the density of ion species $b$ increases to the region where the density of another ion species $a$ increases when $(Z_b - Z_a)(m_a + m_b) + 3/2 (m_a Z_a - m_b Z_b) > 0 $. 
Both heating due to charge incompressibility and the Ettingshausen effect are reversible.

The heat pump produces a temperature gradient, which can be called a \textit{piezothermal effect in magnetized plasma}, in analogy with a similar effect in neutral gas \cite{Geyko2016}. The piezothermal effect is reversible up to the heat conductivity and frictional heating, the same way as piezoelectric effect is reversible up to Ohmic losses.
The resulting temperature differences across the plasma can be comparable to the temperature itself. 
In contrast to neutral gas, however, magnetized multi-ion species plasma features multiple heating mechanisms, so the temperature profile can be tailored by modifying the external potential and plasma parameters.
These properties of magnetized multi-ion-species plasma were successfully observed with the code MITNS \cite{Kolmes2020ii}.

Although the fastest transport timescale in magnetized plasma is associated with ion-ion friction, the plasma features different behavior on longer timescales as it adjusts to a changing field through several distinct collisional processes such as ion-electron friction and ion viscosity. 
This marks a difference with rotating multiple-ion nonneutral plasmas, where similar ion stratification effects are found \cite{ONeil1981, Larson1986, Wineland1993, Imajo1997, Dubin1999, Gruber2005, Gabrielse2010, Andresen2011, Affolter2013, Danielson2015}. For other reasons, ion separation with respect to $m/Z$ occurs in unmagnetized plasma \cite{Kagan2012, Kagan2014}.

The very new effects identified here in ion-ion transport may be important in plasma applications. 
In fusion devices, it is typically advantageous to concentrate fuel ions in the hot, dense core of the plasma and to flush out impurities and fusion products  \cite{Hirshman1981, Redi1991, Braun2010, Ochs2018i, Ochs2018ii, Knapp2019, Gomez2019, Schmit2014}. 
Differential ion transport is even more centrally important for plasma mass filters, which are designed to separate the components of a plasma according to mass \cite{Bonnevier1966, Lehnert1971, Hellsten1977, Krishnan1983, Geva1984, Bittencourt1987, Amoretti2006, Dolgolenko2017, Ochs2017iii, Yuferov2018, Gueroult2018} for a variety of applications \cite{Gueroult2014ii, Gueroult2015, Gueroult2018ii, Gueroult2019}. 
Although some potential applications, such as vortices in Z-pinches \cite{Yu2019}, lie in rotating and compressing high-$\beta$ plasma columns, here for clarity a low-$\beta$, strongly magnetized, two-ion-species plasma slab, with a gravitational force as a proxy for centrifugal and inertial forces, is considered.

\textit{Fluid Model.} Consider a low-$\beta$ plasma slab in a homogeneous magnetic field $\mathbf{B} = B \hat{z}$ with species-dependent external potential $\Phi_s(y, t)$ with $\nabla \Phi_s$ (and all other gradients) in the $\hat{y}$ direction. This plasma can be described by a multiple-fluid model \cite{Braginskii1965, Simakov2003}.
The $\hat{x}$-component of the fluid momentum equation can be written as
\begin{align}
    m_s \frac{d_s v_{sx}}{dt}& = q_s \left( E_x + v_y B \right) - \frac{(\nabla \cdot \pi_s)_x}{n_s} \nonumber \\ & + \sum_{s'} \bigg( \nu_{ss'} m_s (v_{s'x} - v_{sx}) +  \rm{f}^T_{ss',x} \bigg).
\label{eqn:momentum}
\end{align}
Here $d_s / dt = \partial/\partial t + \bv_{s} \cdot \nabla$ is the fluid derivative, $\pi_s$ is the traceless part of the pressure tensor of species $s$, $E_x$ is an inductive electric field, and $\rm{f}^T_{ss'}$ is $\hat{x}$ component of the Nernst (``thermal") friction force density between species $s$ and $s'$ \cite{Hinton, HelanderSigmar}. 
In the limit where $m_s \ll m_{s'}$, this is $\mathbf{f}^T_{ss'} = 3 \nu_{ss'} \hat{b} \times \nabla T_s / 2 \Omega_s$. 
Eq.~(\ref{eqn:momentum}) neglects frictional and inertial terms associated with motion in the $\hat{y}$ direction under the assumption that $v_{sy}$ will be much smaller than the drifts in the diamagnetic direction. Eq.~(\ref{eqn:momentum}) can be rearranged to show that the resulting $\hat{y}$-directed particle flux for species $s$ is 
\begin{align}
\Gamma_{sy} & = - \frac{n_s}{\Omega_s} \bigg[  \sum_{s'} \bigg( \nu_{ss'} (v_{s'x} - v_{sx}) + \frac{1}{m_s} \rm{f}^T_{ss'} \bigg) \nonumber \\
& + \frac{(\nabla \cdot \pi_s)_x}{q_s n_s B} - \frac{\partial v_{sx}}{\partial t} - v_{sy} v_{sx}' \bigg] - \frac{n_s E_x}{B}. \label{eqn:flux}
\end{align}
The density profile will equilibrate on a timescale $L n_s / \Gamma_{sy}$, where $L$ is the length scale of the density gradients. Therefore, the largest terms in Eq.~(\ref{eqn:flux}) correspond to the fastest timescale.
The fastest timescale comes from the ion-ion frictional interactions from the first two terms in Eq.~(\ref{eqn:flux}); $n_i$ reacts to these on a timescale $\tau_{ii'}^{eq} \sim (\nu_{ii'} \rho_{L,i}^2/ L^2 + \nu_{i'i} \rho_{L,i'}^2/ L^2)^{-1}$, where $\rho_{L,i}$ is the Larmor radius of species $i$. The other forces are typically smaller, and their corresponding timescales are proportionally longer. The ion-electron friction has a timescale with $\tau_{ie}^{eq} / \tau_{ii'}^{eq} \sim \max_{i'} Z_{i'}^2 (n_{i'} / n_e) \sqrt{m_{i'}/m_{e}}$. 
Net charge transport across magnetic field lines occurs due to the other terms in Eq.~(\ref{eqn:flux}), which are even smaller.

Consider timescales long enough for the system to balance the ion-ion friction but not long enough for ion-electron or viscous effects to have played any role, which are longer by a factor of $\tau_{ie}^{eq}/\tau_{ii'}^{eq}$, which scales at least as large as $\sqrt{m_i / m_e} .$
From now on, the $\tau_{ii'}^{eq}$ effects are described.
Timescales are the same when centrifugal or inertial forces are applied to a plasma cylinder. As such, all the effects described here can be found in rotating or compressing plasmas, too.

\textit{Charge Incompressibility.} 
When the displacement currents associated with fast changes to the potential can be neglected and all gradients are in the $\hat{y}$ direction, Amp\`ere's Law gives that $j_y = 0$; if the plasma is initially quasineutral, it will remain so. Therefore, importantly, electron motion dictates the evolution of the total ion charge density $\rho_{ci}$, defined as
\begin{gather}
\rho_{ci}(y) \doteq \sum_i q_i n_i(y). \label{eqn:rhoC}
\end{gather}
The continuity equation relates the evolution of the electron density to the electron flux $\Gamma_e$, which can be found from Eq.~(\ref{eqn:flux}). 
If electrons cannot flow through the boundaries of the system, Faraday's Law gives that $E_x \propto \partial B / \partial t \propto \beta $. Then $|\Gamma_e / \Gamma_i| \sim \mathcal{O} (\tau_{ii'}/\tau_{ie}, \beta)$, and in the low-$\beta$ limit total ion charge density $\rho_{ci}$ is conserved on $\tau_{ii'}^{eq}$ timescales. Therefore, charge density (as opposed to mass density) remains ``frozen in" the magnetic field on  $\tau_{ii'}^{eq}$ timescales, even though collisions are important.

Charge incompressibility and ion-ion frictional interaction lead to stratification of ions on the $\tau_{ii'}^{eq}$ timescale. In particular, in the absence of temperature gradients, Eqs.~(\ref{eqn:flux}) and (\ref{eqn:rhoC}) can be combined to find that the densities of ion species $a$ and $b$ are related \cite{Spitzer1952, Taylor1961ii, Braginskii1965, Kolmes2018, Kolmes2020i} 
\begin{gather}
\bigg( n_a e^{\Phi_a / T} \bigg)^{1/Z_a} \propto \bigg( n_{b} e^{\Phi_{b}/T} \bigg)^{1/Z_{b}}. \label{eqn:phiPinch}
\end{gather}
Eq.~(\ref{eqn:phiPinch}) does not include the effects of temperature gradients  \cite{Rutherford1974, Hinton1974, Hinton1976, Wade2000,  HelanderSigmar, Dux2004, Helander2017, Newton2017}, but the relaxation of temperature gradients takes place on timescales similar to $\tau_{ii'}^{eq}$.

\begin{figure*}
    \centering
    \includegraphics[trim = 0cm 0cm 0cm 0cm, width=\linewidth]{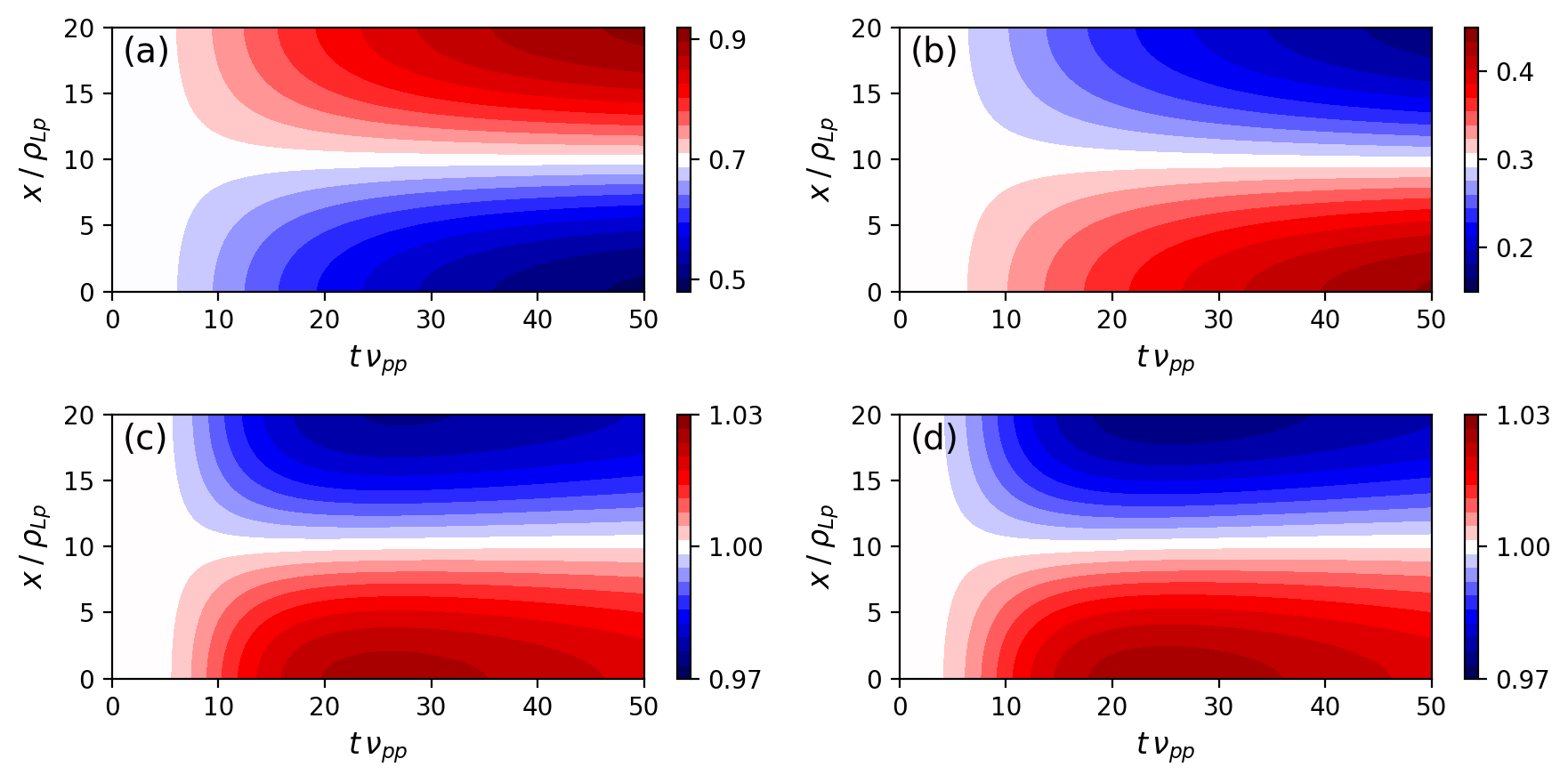}
    \caption{The evolution of: (a) H density; (b) $^{4}$He density; (c) H temperature; (d) $^{4}$He temperature in the MITNS simulation of two-ion-species plasma slab. The net ion density is increasing at the top, but the Ettingshausen effect causes the net heat flow to be from top to bottom. Gravitational potential increases from zero to $m_p g L = 2 T_0$; the initial $\beta$ is $10^{-3}$; the system size is 20 proton gyroradii; and the proton inverse Hall parameter is $\nu_{pp} / \Omega_p \sim 0.05$ (for characteristic parameters). Density and temperature units are arbitrary. The potential $\Phi_s$ is shaped like the segment of a sine curve from minimum to maximum.}
    \label{fig:EttingshausenHHe}
\end{figure*}

\textit{Heating Mechanisms.} 
After the potential is applied, stratification of ions leads to the release of a fraction of the potential energy. The change of electric field energy, magnetic field energy, and energy of drift motion are $\mathcal{O} (\beta^2), \mathcal{O} (\beta), \mathcal{O} ((\rho_{L,i}/L)^2)$ smaller than heating, respectively. Therefore, in low-$\beta$ plasma, the released potential energy goes into heating.
A magnetized plasma slab has multiple ways to transform released potential energy into heat, as can be seen from the temperature evolution equation in a multispecies plasma \cite{HazeltineWaelbroeck,Braginskii1965}: 
\begin{align}
\frac{3}{2} n_s \frac{d_s T_s}{dt} - T_s \frac{d_s n_s}{dt} = - \nabla \cdot \left(\bq_s^T + \bq_s^u \right) + Q_s . \label{eqn:temperature}
\end{align}

Here $d_s / dt = \partial/\partial t + \bv_{s} \cdot \nabla$ is the fluid derivative, $\bq_s^T$ is a component of heat flux of species $s$ that is $\propto \nabla T_s$, and $Q_s = \sum_{s'} \left[ Q_{ss'} -  (\bv_s - \bv_{s'}) \cdot \mathbf{R}_{ss'} \cdot m_{s'}/(m_s + m_{s'}) \right]$ is the energy gained by species $s$ through collisions with all  species $s'$ due to temperature equilibration $Q_{ss'}$ and due to frictional heating ($\mathbf{R}_{ss'}$ is total friction force between species $s$ and $s'$). However, frictional heating is negligible as long as characteristic rate of change of the potential $|\dot{\Phi}_s/\Phi_s|$ is comparable to $(\tau_{ii'}^{eq})^{-1}$. $\bq_s^u$ is the heat flux due to the difference of flow velocities (Ettingshausen effect). It is an Onsager-symmetric effect to the thermal friction force $f_{ss'}^T$. The usual expression for heat flux in multispecies plasma \cite{Hinton} is not affected by low-flow ordering corrections \cite{Simakov2003}. Viscous heating $-\pi_s \colon \nabla \bv_s$ is negligible, so it is omitted in Eq.~(\ref{eqn:temperature}). 
Following \cite{Hinton}, $q_{sy}^u$ is
\begin{gather}
q_{sy}^u = \frac{n_s T_s}{\Omega_s} \sum_{s'} \frac{3}{2} \frac{\nu_{ss'}}{1 + m_s/m_{s'}} \left( v_{s'x} - v_{sx} \right).
\label{eqn:fluxEttingshausen}
\end{gather}
Note that the largest contribution to $q_{sy}^u$ is due to the difference in ion drift velocities. It is proportional to the ion cross-field particle flux, $q_{sy}^u \sim T_s \Gamma_{sy}$. In a multi-ion species plasma, the heating due to the Ettingshausen effect $\nabla \cdot \bq_s^u$ is proportional to the heating due to the ion density change $T_s \partial n_s / \partial t$.

\textit{Piezothermal Effect.}
In a neutral gas, when gravity is increased, particles at the bottom of a slab see a falling ceiling, so they are heated more than particles at the top, which see a falling floor. Therefore, a temperature gradient develops with $T_{max}$ at the bottom. Since this effect has been referred to in neutral gas as a piezothermal effect, we adopt the term \textit{piezothermal effect in low-$\beta$ magnetized plasma} to describe the temperature gradient formation as the external potential $\Phi$ is changed.

To calculate how large these effects can be, suppose that the initial density and temperature are uniform, and $\Phi_s = 0$. 
Let $\Delta \Phi$ be the difference between $\Phi_a/Z_a - \Phi_b/Z_b$ at the top of the slab and at the bottom after the potential has changed. 
When $\Delta \Phi/ T \lesssim 1$, heating due to charge incompressibility and the Ettingshausen effect cause the piezothermal effect, as convective terms in Eq.~(\ref{eqn:temperature}) are $\propto (\Delta \Phi/ T)^2$. On the fastest timescale, ion-ion friction is the dominant cross-field transport mechanism, so in two-ion-species plasma, neglecting thermal friction, Eqs.~(\ref{eqn:flux}) and (\ref{eqn:fluxEttingshausen}) can be combined to get 
\begin{align}
- \nabla \cdot \bq_a^u = - \frac{3}{2} \frac{m_b}{m_a + m_b} T_a \frac{\partial n_a}{\partial t}.
\label{eqn:compressionEttingshausen}
\end{align} 
Suppose ion species $a$ and $b$ are in local temperature equilibrium and have $\Gamma_a = \Gamma_b = 0$. 
Then, neglecting diffusive heat flux $\bq^T$, combine Eqs.~(\ref{eqn:flux}), (\ref{eqn:rhoC}), (\ref{eqn:temperature}), and (\ref{eqn:compressionEttingshausen}) to get
\begin{align}
\frac{\nabla T}{T} & =  C \left(\frac{\nabla \Phi_b}{Z_b T} - \frac{\nabla \Phi_a}{Z_a T} \right) \left[\frac{3n}{2}\left(\frac{1}{Z_a^2 n_a} + \frac{1}{Z_b^2 n_b}\right) + C^2\right]^{-1}; \nonumber\\  C & = \left[\left(\frac{1}{Z_a} - \frac{1}{Z_b}\right) + \frac{3}{2}\frac{m_a/Z_b - m_b/Z_a}{m_a + m_b}\right].
\label{eqn:temperatureGradient}
\end{align}
Here $n$ is total density of species that are in local temperature equilibrium ($n = n_a + n_b$ if $\nu_{ie}^{-1} \gg \tau_{ii'}^{eq}$, $n = n_a + n_b + n_e$ if $\nu_{ie}^{-1} \ll \tau_{ii'}^{eq}$).
Eq.~(\ref{eqn:temperatureGradient}) predicts which ion species are heated and which are cooled. If $C > 0$, it leads to heating in the region where the concentration of species $a$ increases, and cooling where it decreases. Therefore, heat is transferred from species $b$ to species $a$. For example, in a plasma of tritium and ${}^4$He, helium goes to the top of the slab and is heated, while tritium goes to the bottom and is cooled.
If $C < 0$, the roles of species $a$ and $b$ are reversed. Another example of this is shown in Figure~\ref{fig:EttingshausenHHe}.
This effect can be observed only in multi-ion species plasma.

\begin{figure}
\centering
\includegraphics[trim={0.2cm 0.2cm 0.2cm 0.2cm},clip,width=0.9\linewidth]{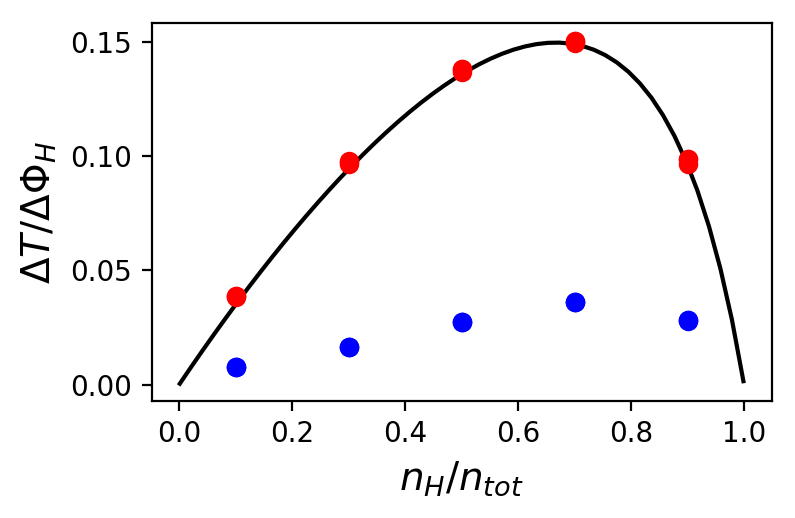}
\caption{Maximum spatial variation of temperature $\Delta T$ versus hydrogen fraction in H-He plasma. The simulation results: with diffusive heat flux $\bq^T$ (blue) and without $\bq^T$ (red). The prediction of Eq.~(\ref{eqn:temperatureGradient}) is the black line. Points represent values of $m_p g L / T_0$ ($0.01, 0.1, 1$), all overlapping.}
\label{fig:maximumTemperatureDifference}
\end{figure}

The size of the effect can be significant. Indeed, suppose there is a $r = 0.1~cm$ cylindrical vortex of H-He plasma with the following parameters: $T_i = 100~eV$, $n_e = 10^{16}~cm^{-3}$, $2 n_{He} = n_{H}$, which is immersed into axial magnetic field, $B = 10~T$. Plasma with somewhat similar parameters occurs in magnetized target fusion \cite{Intrator2004i,Intrator2004ii} with a somewhat lower and differently configured magnetic field. Suppose this vortex is solid-body rotating at angular velocity $\omega = 6.2 \cdot 10^7~s^{-1}$. Such a vortex has $\beta = 7\cdot10^{-3}$, $\nu_{pp}/\Omega_p = 5.7\cdot 10^{-3}$, $\rho_p / r = 0.1$, $m_p \omega^2 r^2 / 2 T = 0.2$.  Then, if this vortex is subjected to radial metric compression such that radius is decreased by a factor of $30$ over the compression time $\tau_{c} = 3\cdot 10^{-6}~s$ and if $\gamma = 5/3$, the edge of the vortex is hotter than the center by $12\%$ in the absence of heat diffusion (i.e. temperature at the edge is $9.9~keV$ and temperature at the center is $8.8~keV$) when the size of the temperature difference is the largest. In this estimate, some effects such as decrease of adiabatic index $\gamma$ due to rotation \cite{Geyko2013,Geyko2017} and constraint on angular momentum transport are ignored. However, these effects are only going to increase the size of temperature difference.
Also, plasma is going to be more magnetized as it is compressed.

Although heat flux $\bq^T$ is not included in Eq.~(\ref{eqn:temperatureGradient}), it is incorporated in the code MITNS (Figure~\ref{fig:maximumTemperatureDifference}). The fact that, absent heat flux $\bq^T$, the analytic and numerical solutions agree, gives confidence in both. Note that, heat conduction does not change the location of extrema of temperature profiles, and only relaxes temperature gradients by an $\mathcal{O}(1)$ factor because the timescale of ion-ion equilibration, $\tau_{ii'}^{eq}$, is similar to the timescale of cross-field heat transport, $\tau_{th}^{eq} \sim \tau_{ii'}^{eq}$.
Note that, due to the similarity of $\tau_{th}^{eq}$ and $\tau_{ii'}^{eq}$ timescales, the piezothermal effect in low-$\beta$ magnetized plasma is not reversible.
This is different from the neutral gas case, where the heat transport timescale is much larger than force equilibration timescale so that the piezothermal effect is reversible.

When $\Delta \Phi /T \gtrsim 1$, the shape of the temperature profile can be more complex, since rearrangement heating can play a prominent role in this case. In particular, the temperature profile can be tailored to peak at the separation layer or at the region occupied by one of the ion species as predicted by Eq.~(\ref{eqn:temperatureGradient}).
Also for $\Delta \Phi /T \gtrsim 1$, spatial variation of thermal conductivity distorts the shape of the temperature profile while not altering the locations of the temperature extrema.

\textit{Summary.} 
A new heat pump effect in magnetized plasma has been identified. To compare analytic to computational descriptions of this effect, a time-varying gravitational field is used as a proxy for similar forces, like the centrifugal force in a rotating system or the inertial force in a compressing system. The behavior of a low-$\beta$ magnetized net neutral plasma in a perpendicular gravitational field is then shown to be markedly different from that of either a nonneutral magnetized plasma, a neutral gas, or an unmagnetized plasma. 
Over timescales $\sim \tau_{ii'}^{eq}$, the ions assort themselves based on $m/Z$ ratio. If $\Delta \Phi / T$ is large, ions can be arbitrarily well-stratified.
On these timescales, magnetized multi-ion species plasma is essentially a new kind of substance; its behavior appears to feature traditional buoyancy, but with a constraint tied to charge density rather than number density. The ion separation is an initial condition-dependent, $\tau_{ii'}^{eq}$-timescale effect in a quasineutral low-$\beta$ plasma, in contrast to rotating nonneutral plasmas where similar ion stratification is a state of thermodynamic equilibrium \cite{ONeil1981}.

A new and remarkable phenomenon is that the ion stratification on $\tau_{ii'}^{eq}$ timescales leads to a significant heat pump via charge incompressibility and the Ettingshausen effect. The resulting temperature gradients are predicted by Eq.~(\ref{eqn:temperatureGradient}) up to the corrections due to the diffusive heat flux $\bq^T$. If $\Delta \Phi / T$ is large, the size of possible temperature differences across the magnetized plasma can be substantial, namely $(T_{max} - T_{min})/ T_{min} = \mathcal{O}(1)$. 

The authors thank R. Gueroult, S. Davidovits, and M. Kunz for useful conversations. This work was supported by Cornell NNSA 83228- 10966 [Prime No. DOE (NNSA) DE-NA0003764] and by NSF-PHY-1805316.

\providecommand{\noopsort}[1]{}\providecommand{\singleletter}[1]{#1}%

\end{document}